\documentclass[amstex,aps,preprint,showpacs]{revtex4}

\usepackage{graphicx}
\usepackage{amsmath}
\usepackage{bm}
\usepackage{amsfonts}
\usepackage{amssymb}

\begin{document}

\title{Entanglement in the classical limit: quantum correlations from
classical probabilities}

\begin{abstract}
We investigate entanglement for a composite closed system endowed
with a scaling property allowing to keep the
dynamics invariant while the effective Planck constant $\hslash_{\mathrm{eff}}$
of the system is varied. Entanglement increases as $\hslash_{\mathrm{eff}}\rightarrow0$.
Moreover for sufficiently low $\hslash_{\mathrm{eff}}$ the evolution
of the quantum correlations, encapsulated for example in the quantum
discord, can be obtained from the mutual information of the corresponding
\emph{classical} system. We show this behavior is due to the local
suppression of path interferences in the interaction that generates the entanglement. This
behavior should be generic for quantum systems in the classical limit.
\end{abstract}

\author{A. Matzkin}
\affiliation{LPTM (CNRS Unit\'{e} 8089), Universit\'{e} de Cergy-Pontoise, 95302
Cergy-Pontoise cedex, France}
\pacs{03.67.Mn,05.45.Mt,03.65.Sq}
\maketitle

Entanglement is a distinctive feature of quantum mechanics, ``\emph{the one
that enforces its entire departure from classical lines of thought}''
\cite{schroedinger}. Its understanding has tremendously progressed in the
last decade, due essentially to a vast amount of work regarding the
construction and properties of entangled qubits in view of possible
applications in quantum information \cite{horodecki}. In a more general
context, \emph{any} dynamical interaction between quantum particles leads to
entanglement, that stands as a formidable obstacle to account for the
emergence of classical word. Explaining the unobservability of entanglement
in the classical limit is one of the aims of the decoherence programme \cite%
{zurek}.

Somewhat more modestly, several recent works \cite{refQC} have studied in
semiclassical systems the link between the generation of entanglement and
the dynamics of the corresponding classical system, including in experimental
realizations \cite{exp}. The numerical and analytical results obtained so far
indicate that the entanglement dynamics in
quantum systems having a classically chaotic counterpart sharply differs
from those whose classical
counterpart is regular, though this difference is dependent on the specificities of the considered systems (types and
strengths of the coupling, choice of initial states, etc.). It has been argued
\cite{argued} that a proper understanding of the connection between the
classical dynamical regime and entanglement hinges on employing systems in
which the same physical process generates the dynamics in the classical
system and entanglement in its quantum counterpart.

An intriguing question studied in this paper concerns the behavior of
entanglement in these systems when the typical actions of the system grow with respect to $%
\hslash$. Then the size of the Hilbert space increases and the
quantum-classical correspondence improves. Moreover if the system dynamics
can be kept invariant while the actions increase, an effective Planck
constant $\hslash _{\mathrm{eff}}$ can be defined and entanglement can be
studied as $\hslash_{\mathrm{eff}}\rightarrow0$. We will see that
entanglement indeed increases with the size of the Hilbert space in
agreement with previous findings on entangled Bose-Einstein condensates \cite{angelo}.
Maybe more surprisingly for sufficiently low $\hslash_{%
\mathrm{eff}}$ the evolution of the entanglement measure is given by
probabilities obtained from the classical dynamical evolution, irrespective
of the dynamical regime. A consequence discussed below is whether in the $%
\hslash_{\mathrm{eff}}\rightarrow0$ limit the quantum information
encoded in the pure state density matrix becomes indiscernible from
the classical information contained in a mixed density matrix yielding the same reduced
dynamics.

Let us consider bipartite entanglement generated by repeated inelastic
scattering of two particles. To set the model, let us assume a
light structureless particle and a heavy rotating particle, modeled by a
symmetric top with angular momentum $N$ and energy $E_{N}= N(N+1)/2I$, $I$ denoting
the moment of inertia.
The scattering potential is taken to be a contact interaction so that the
light incoming particle receives a kick when it hits the rotating top. The
conservation of the total angular momentum $\mathbf{T}=\mathbf{N}+\mathbf{J}$
where $J$ is the light particle angular momentum, imposes that after the collision
the rotating top is left with an angular momentum $N^{\prime}$ obeying $%
T-J\leq N^{\prime}\leq T+J$ where we have assumed $J^{\prime}=J$. The probability of
the transition $N\rightarrow N^{\prime}$ is obtained from the scattering
matrix elements $\left\vert S_{NN^{\prime}}\right\vert ^{2}$. Finally to account for repeated scattering we need an
attractive long-range field between both particles: we will assume the
particles have opposite electric charge. Note that this model can be seen as
a two-particle extension of the standard kicked top well-known in quantum chaos \cite{exp,haake}.

\begin{figure}[tb]
\includegraphics[height=4.5cm]{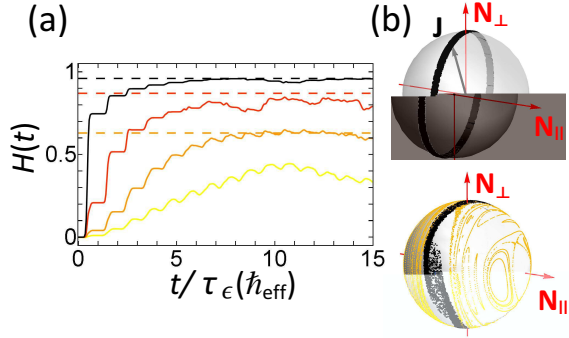}
\caption{(a) The entanglement rate as a function
 of the number of kicks is given for quantum systems characterized by $k=0.25$ and
$\hbar_{\mathrm{eff}}=0.5,0.25,0.1,0.01$ (from bottom to top). The dashed lines
result from applying the scaling relation (\ref{simplef}) to the 3 upper curves.
(b) Top: The black ring encircling the $\mathbf{N}$ axis is the initial classical distribution of $\mathbf{J}$
with $J=10$ (corresponding to the quantum case with $\hbar_{\mathrm{eff}}=0.1$), centered on $N_0=T=50$. Bottom: The
distribution at $t=10$ kicks (black dots), showing a spread. The surface of section
for  $k=0.25$ is shown in yellow (light gray).}
\label{figT1}
\end{figure}

Starting from an initial product state $\left\vert \psi _{0}\right\rangle
\equiv \left\vert F_{0}^{-}(\bar{\epsilon}_{0})\right\rangle \left\vert
N_{0}\right\rangle $ where $\left\vert F_{0}^{-}\right\rangle $ depicts an
incoming wavepacket of the light particle with mean energy $\bar{\epsilon}%
_{0}$ traveling towards the rotating top in state $\left\vert
N_{0}\right\rangle ,$ entanglement is generated as soon as the first
collision takes place. The outgoing wavefunction is then given by the
superposition $\sum_{N}S_{NN_{0}}\left\vert F^{+}(\epsilon
_{N})\right\rangle \left\vert N\right\rangle $ where the dependence of $%
\epsilon $ on $N$ is due to the conservation of energy, $\epsilon
_{N}=E-E_{N}$, with $E=\bar{\epsilon}_{0}+E_{N_{0}}$ being the total energy.
The scattered wavepackets are later turned back by the attractive field and
are treated as newly incoming waves. The pure state density matrix%
\begin{equation}
\rho (t)=U(t,t_{0})\left\vert \psi _{0}\right\rangle \left\langle \psi
_{0}\right\vert U^{+}(t,t_{0})  \label{n13}
\end{equation}%
is obtained by writing the evolution operator $U$ in terms of the scattering
eigenstates of the Hamiltonian
\begin{eqnarray}
\left\vert \psi (E)\right\rangle &=&\sum_{N}Z_{N}^{-}(E)\left\vert
F^{-}(\epsilon _{N})\right\rangle \left\vert N\right\rangle
+\sum_{NN^{\prime }}  \notag \\
&&Z_{N}^{-}(E)S_{N^{\prime }N}\left\vert F^{+}(\epsilon _{N^{\prime
}})\right\rangle \left\vert N^{\prime }\right\rangle  \label{n15}
\end{eqnarray}%
where $Z_{N}^{-}$ are coefficients obtained by applying the asymptotic
boundary conditions%
. The maximal number of entangled states is given by the
number of scattering channels $n=2J+1$. The amount of entanglement will be
estimated through the entropy of the reduced density matrix. We will employ
the linearized form%
\begin{equation}
H(t)=\frac{n}{n-1}\left( 1-\mathrm{Tr}_{N}\rho _{N}^{2}(t)\right)  \label{17}
\end{equation}%
that becomes a good
approximation for large $n$. $\rho _{N}(t)$ is the reduced density matrix
obtained by tracing over the light particle's degrees of freedom. Note $H=1$
for a maximally entangled state. For convenience we set from now on $%
t_{0}=\tau _{\epsilon }/2$ where $\tau _{\epsilon }$ is the period of the
mean energy orbit; then the collision times are $t=q\tau _{\epsilon }$ with $%
q$ being an integer.

The classical version of the model can be formally obtained
by employing the semiclassical link \cite{lombardi04} between the deflection angle $\phi $
produced by the torsional motion and the eigen-phaseshifts $\delta $ of the $%
S$-matrix: in the top's reference frame, each kick rotates $\mathbf{J}$ by an
angle $\phi =kJ_{\bot }/J=\partial \delta /\partial J_{\bot }$ where $%
J_{\bot }$ is the projection of $\mathbf{J}$ on the unit axis $\mathbf{\hat{N%
}}_{\bot }$ perpendicular to $\mathbf{N}$. $k$ is the strength of the
kick; a given $k$ corresponds, via the semiclassical relation, to a given  $S$-matrix, i.e.
$S_{NN^\prime}=S_{NN^\prime}(k)$. The classical orbit of the light particle
between two scattering events induces a rotation of $\mathbf{J}$ around $%
\mathbf{N}$ by an angle $2\pi \tau _{\epsilon }/\tau _{N}$ where $\tau _{N}$
is the top rotation period. A surface of section is obtained by plotting the
position of $\mathbf{J}$ after each kick (see Figs 1(b) and 3).
The crucial observation is that the surface of section only depends on $k$
and on $\tau _{\epsilon }/\tau _{N}$: $N,J$ and $T$ (which are action
variables) can be increased at will, say by division by $\hslash _{\mathrm{%
eff}},$ but the dynamical map stays constant provided $E$ and $I$ are adjusted
accordingly. For a long-range central field this also
implies dividing the radial action $W_{r}$ of the light particle by the same
constant $\hslash _{\mathrm{eff}}.$ Hence multiplying $N,J,T$ and $W_{r}$
by the common factor $1/\hslash _{\mathrm{eff}}^{{}}\gg 1$ is tantamount to
studying the limit $\hslash \rightarrow 0$ without modifying the underlying
dynamics.\ Note that the number of entangled states $n$ also scales with $%
1/\hslash _{\mathrm{eff}}.$

Fig.\ 1(a) displays the entanglement evolution for the quantum two-particle
kicked top with $k=0.25$ for different values of $\hslash _{\mathrm{eff}}$
(we employ atomic units and set $\hbar = 1$). The light particle's
initial distribution is a Gaussian wavepacket localized far from the
symmetric top with its mean initial momentum directed towards it. The
entanglement increases dramatically as $\hslash _{\mathrm{eff}}$ decreases,
despite the fact that the initial state takes a smaller relative area on the
sphere.\ To first order, this is a consequence of the similarity
transformation: on the one hand $\rho _{N}(t)$ is by definition a convex
combination of projectors $\left\vert N\right\rangle \left\langle
N\right\vert $, and on the other hand in the semiclassical approximation the
projection of $\rho _{N}(t)$ on the unit sphere (at kick times $t=q\tau
_{\epsilon })$ covers the same area irrespective of $\hslash _{\mathrm{eff}}$%
. Let $m$ be the number of projectors $\left\vert N\right\rangle
\left\langle N\right\vert $ (out of total of $n$) projecting in this area
for some $\hslash _{\mathrm{eff}}$ and $m^{\prime }$ that number for another
choice of $\hslash _{\mathrm{eff}}^{^{\prime }}<\hslash _{\mathrm{eff}}$.
Then $m/n=m^{\prime }/n^{\prime }$ from which it follows that for situations
corresponding to the maximal entanglement (uniform distribution in that
region) there is a simple scaling relation for the purity $1-H$ yielding%
\begin{equation}
H^{\prime }(t)=1-\frac{\hslash _{\mathrm{eff}}^{\prime }}{\hslash _{\mathrm{%
eff}}}\left( 1-H(t)\right) . \label{simplef}
\end{equation}%
As expected entanglement increases with the number of available quantum states.

\begin{figure}[tb]
\includegraphics[height=5cm]{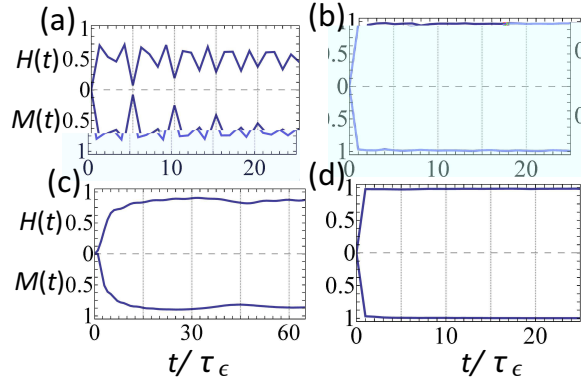}
\caption{Each panel shows the entanglement evolution as computed from the entropy $H(t)$ (top curve),
along with the mutual information $M(t)$ of the \emph{classical} counterpart (upside-down curve), where $t$
is computed at the discrete kick times $t=q\tau _{\epsilon }$. The parameters are
$J=100, T=500$ (corresponding to $\hbar_{\mathrm{eff}}=0.01$ for the quantum system) and
(a): $k=1, N_0=402$, (b): $k=1, N_0=430$, (c): $k=0.01, N_0=498$, (d): $k=10, N_0=460$.}
\label{figT2}
\end{figure}

The classical evolution analogue of the quantum problem leading to Fig.\ 1(a)
is obtained by taking an initial uniform distribution of angular momentum $%
N_{0}-\frac{1}{2n}<N<N_{0}+\frac{1}{2n}$, i.e. we cut the sphere along the $%
\mathbf{N}$ axis into $n$ slices of width $1/n$. Each slice is centered so that the
projection $J_{\parallel }$ on the
$\mathbf{\hat{N}}_{\parallel}$ axis matches integer values of $N$. Thus
the initial classical distribution corresponding to
$\left\vert N_{0}\right\rangle \left\langle N_{0}\right\vert $, is represented on the sphere by the ring centered at
$J_{\parallel }(N_{0})=\mathbf{T}\cdot \mathbf{\hat{N}}_{\parallel%
}-N_{0}$.
The light particle initial distribution is the same
Gaussian employed in the quantum problem (the role of this distribution is
to give a statistical weight, depending on the initial energy of the light
particle, to each $\mathbf{J}$ lying in the initial ring). We compute
numerically the evolution (torsion during the kick and rotation during the
orbital excursion) for each $\mathbf{J}$ of the initial distribution from
kick to kick. The classical probability $p_{N}^{cl}(t=q\tau _{\epsilon })$ of
finding the top with an angular momentum in the interval $\Delta _{N}=[N-%
\frac{1}{2n},N+\frac{1}{2n}]$ after $q$ kicks is obtained by counting the
vectors $\mathbf{J}$ whose projection falls in the corresponding interval.
The probabilities after the first kick are given by the transition
probability $P^{cl}(\Delta _{N_{0}}\rightarrow
\Delta _{N})$.
The classical
probabilities after $q$ kicks obey the recurrence relation%
\begin{equation}
p_{N}^{cl}(q\tau _{\epsilon })=\sum_{N^{\prime }}P^{cl}(\Delta _{N^{\prime
}}\rightarrow \Delta _{N})p_{N^{\prime }}^{cl}\left[ \left( q-1\right) \tau
_{\epsilon }\right] .  \label{n6}
\end{equation}%
From these probabilities one can define the quantity
\begin{equation}
M(q\tau _{\epsilon })=\frac{n}{n-1}\left( 1-\sum_{N}\left[ p_{N}^{cl}(q\tau
_{\epsilon })\right] ^{2}\right)  \label{20}
\end{equation}%
that can be understood equivalently as the linear entropy for the total
system or as the (linearized) classical mutual information \footnote{%
This is the case provided the classical distribution is considered in
configuration space, the integral over $N$ being separated as $%
\int_{T-J}^{T+J}..dN=\sum_{\Delta _{N}}\int_{\Delta _{N}}..dN$}, quantifying
the amount of mixing among the $n$ slices induced by the kicks.

Fig.\ 2(a)-(d) shows in the top panel the entanglement evolution as given by $%
H(t=q\tau _{\epsilon })$ for a choice of system parameters (coupling $%
k $ and initial state) all corresponding to $\hslash _{\mathrm{eff}}\approx
1/100$, two orders of magnitude smaller than the hard quantum case $\hslash
_{\mathrm{eff}}\approx 1$ (which is the typical value for qubits), but still
considerably larger than typical values characterizing classical actions.
The bottom panel in each plot shows
$M(q\tau _{\epsilon })$ obtained from the classical probabilities through Eq.
(\ref{20}). The good agreement between $H(t)$ and the
time-dependent classical probabilities holds for classically chaotic and
regular regimes alike as can be inferred from Fig. 3, displaying the
corresponding surfaces of section along
with the classical distributions whose spread along the $\mathbf{N}_\parallel$
axis accounts for the entanglement evolution.

Although quantifying entanglement by means of classical probabilities
might appear surprising at first sight, we expect this behavior to be
generic for semiclassical systems that undergo a loss of phase coherence.\
This is indeed the first ingredient by which the classical $M(q\tau
_{\epsilon })$ can account for $H(t)$.\ The second ingredient is the
semiclassical approximation itself that allows to express operator matrix
elements in terms of classical quantities (the action and the density of
paths). For the system under consideration we start by writing Eq. (\ref{n15}%
) in the form $\left\vert \psi (E)\right\rangle =\sum_{N}B_{N}(E)\left\vert
F(\epsilon _{N})\right\rangle \left\vert N\right\rangle $ where $\left\vert
F\right\rangle $ is a standing wave obtained by combining the $\left\vert
F^{\pm }\right\rangle $ and $B_{N}(E)\equiv \sum_{N^{\prime }}S_{NN^{\prime
}}Z_{N^{\prime }}^{-}(E)e^{i(W_{r}^{po}(\epsilon _{N})-\pi )/2};W_{r}^{po}$
is the radial action of the classical periodic orbit in the attractive field.
Then Eq. (\ref{n13}) takes the form \footnote{%
We assume that the standing waves dependence on the energy is weak and take
the average energy $\bar{\epsilon}_{N}$ within each channel, an assumption
that only holds for radial positions around the classical outer turning
point and thus at times $\tau _{\epsilon }(q+1/2)$.}
\begin{eqnarray}
\rho (t &=&q\tau _{\epsilon }
)=\sum_{NN^{\prime
}}\left\vert N\right\rangle \left\langle N^{\prime }\right\vert
e^{-i(E_{N}-E_{N^{\prime }})t}  \notag \\
&&\beta _{N}(t)\beta _{N^{\prime }}^{\ast }(t)\left\vert F(\bar{\epsilon}%
_{N})\right\rangle \left\langle F(\bar{\epsilon}_{N^{\prime }})\right\vert
\label{25}
\end{eqnarray}%
with
\begin{eqnarray}
\beta _{N}(t) &=&\sum_{N^{\prime }}S_{NN^{\prime }}\left[ \sum_{E}e^{-i%
\epsilon _{N}t}Z_{N^{\prime }}^{-}(E)\left\langle \psi (E)\right\vert \left.
\psi _{0}\right\rangle \right]  \notag \\
&&e^{i(W_{r}^{po}(\epsilon _{N})-\pi )/2}  \label{101}
\end{eqnarray}%
(keep in mind $\epsilon _{N}=E-E_{N}$ when taking the sum). The reduced
density matrix is readily derived as $\rho _{N}(t=q\tau _{\epsilon
})=\sum_{N}\left\vert N\right\rangle \left\langle N\right\vert p_{N}(t)$ with%
\begin{equation}
p_{N}(t=q\tau _{\epsilon })=\left\vert \beta _{N}(t+\frac{\tau _{\epsilon }}{%
2})\right\vert ^{2}=\left\vert \sum_{N^{\prime }}S_{NN^{\prime }}\zeta
_{N^{\prime }}^{-}(t)\right\vert ^{2}.  \label{105}
\end{equation}%
$\zeta _{N^{\prime }}^{-}(t)$ is defined by the term between $[..]$ in Eq. (%
\ref{101}) (i.e. by excluding the phase term in the sum). $\left\vert \zeta
_{N^{\prime }}^{-}(t)\right\vert ^{2}$ represents the probability on the
incoming channel $N^{\prime }$ just before the collision, whereas $%
\left\vert \sum_{N^{\prime }}S_{NN^{\prime }}\zeta _{N^{\prime
}}^{-}\right\vert ^{2}$ is the weight of the outgoing wave right after the
collision ($t=q\tau _{\epsilon }$); in the semiclassical limit this is the
same as the weight $\left\vert \beta _{N}\right\vert ^{2}$ at the apogee
half a period later \footnote{%
Formally the $\tau _{\epsilon }/2$ time shift is obtained by applying the
radial boundary conditions to Eq. (\ref{101}) and then expanding $%
W_{r}^{po}(\epsilon _{N})$ around $\bar{\epsilon}_{N}$ in $\left\vert \beta
_{N}\right\vert ^{2}.$}. It follows that $\zeta _{N}^{-}(q\tau _{\epsilon
})=p_{N}\left[ \left( q-1\right) \tau _{\epsilon }\right] .$ Finally we
recall \cite{lombardi04} that in the semiclassical regime the $S$-matrix elements are given to
first order in $\hslash $ by%
\begin{equation}
S_{NN^{\prime }}=\mathcal{A}_{NN^{\prime }}e^{i\mathcal{S}_{NN^{\prime
}}/\hslash }\text{ with }\left\vert \mathcal{A}_{NN^{\prime }}\right\vert
^{2}=P^{cl}(\Delta _{N^{\prime }}\rightarrow \Delta _{N})  \label{110}
\end{equation}%
and $\mathcal{S}_{NN^{\prime }}$ is the classical action (the boundary
conditions for the conjugate momenta obey $p_{\theta }(t\rightarrow
-\infty )=N^{\prime }$ $p_{\theta }(t\rightarrow +\infty )=N$). As $\mathcal{S%
}_{NN^{\prime }}/\hslash \rightarrow \infty $ the phase terms $\exp i(%
\mathcal{S}_{NN_{1}}-\mathcal{S}_{NN_{2}})/\hslash $ in Eq. (\ref{105})
oscillate wildly, while the amplitudes $\mathcal{A}_{NN^{\prime }}$ are of
the same order of magnitude.\ As a result these off-diagonal terms are
suppressed, so we may keep only the terms with $N_{1}=N_{2}$.\ Eq. (\ref{105}%
) becomes%
\begin{equation}
p_{N}(t=q\tau _{\epsilon })=\sum_{N^{\prime }}P^{cl}(\Delta _{N^{\prime
}}\rightarrow \Delta _{N})p_{N^{\prime }}\left[ \left( q-1\right) \tau
_{\epsilon }\right] .  \label{115}
\end{equation}%
Comparing with Eq. (\ref{n6}) and given that the initial conditions are
identical in the quantum and classical problems, we see that provided the
approximations employed hold, the entanglement entropy $H(t=q\tau _{\epsilon
})$ becomes identical to the mutual information $M(q\tau _{\epsilon })$ of
the corresponding classical system given by Eq. (\ref{20}), thereby
explaining the numerical results displayed in Fig.\ 2.

A remarkable consequence of the present results concerns the classical
values taken by quantifiers of quantum correlations.\ For example the
quantum discord $D(\rho )$ \cite{zurek01} widely employed in the context of
qubit density matrices, measures the quantum information that can only be
extracted by joint measurements on both subsystems. $D(\rho )$ vanishes if
the state has only classical correlations. Here $D(\rho )$ is simply given
by $H(t),$ hence by the classical mutual information $M(t)$. Put
differently, the quantum information contained in the entangled state --
which would be the information gained by an observer making a measurement (for
example measuring the light particle's energy $\epsilon_{N_{m}}$ projects the top to the rotational state $%
\left\vert N_{m}\right\rangle $) -- is given by the ignorance spread arising
from the dynamical evolution of the corresponding classical system.\

\begin{figure}[tb]
\includegraphics[height=4cm]{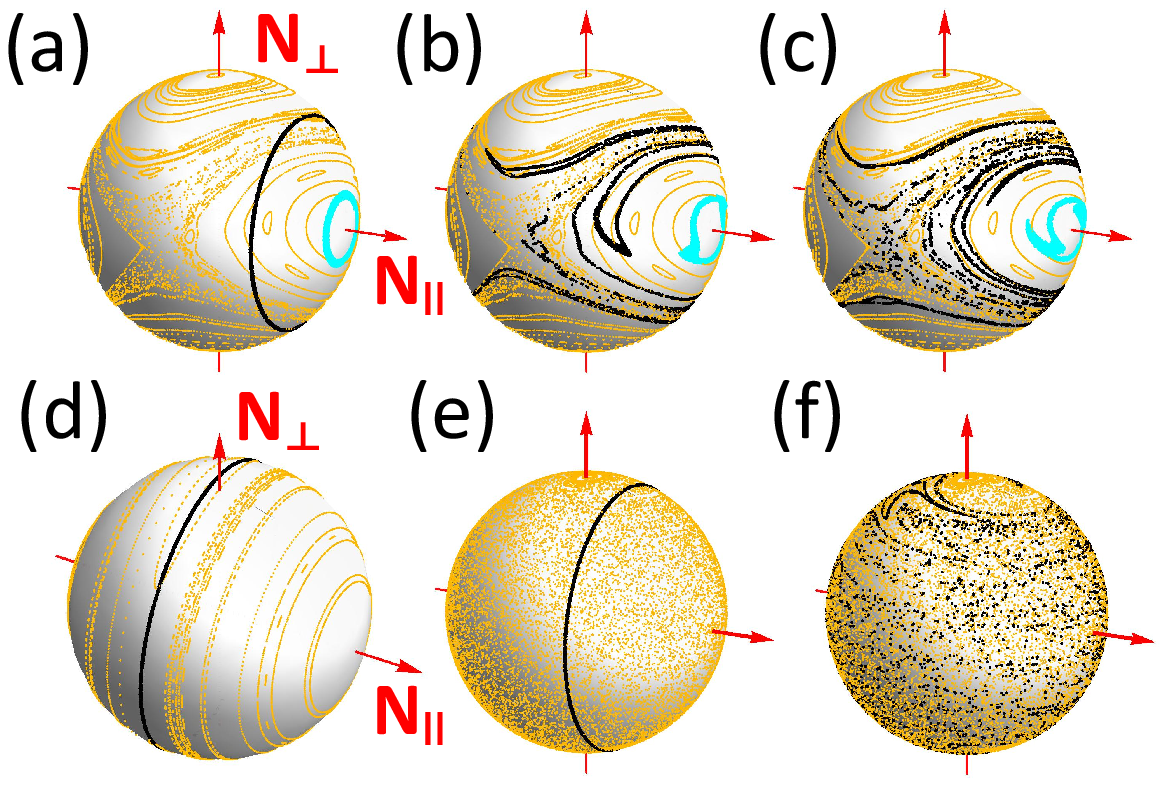}
\caption{Classical evolution and surfaces of section for the parameters corresponding
to the plots of Fig. 2. (a)-(c): for
$k=1$ (mixed phase-space), (a) gives the initial distributions for $N_0=402$ (blue [grey] ring) and
$N_0=430$ (black ring); (b) displays the
distributions after 10 kicks, (c) at $t=25$. (d): position of the initial distribution defined by $N_0=498$
for $k=0.01$ (regular dynamics). (e)-(f): Distributions for $k=10$ (mostly chaotic phase-space) and
$N_0=460$ at $t=0$ [(e)] and at $t=5$ kicks [(f)].
}
\label{figT3}
\end{figure}

The conjunction of ubiquitous entanglement as the classical limit is
approached, and the role played by classical probabilities to account for
quantum correlations in that limit allows us to speculate whether from a
\emph{statistical} perspective, entanglement may be converted within the closed
system into classical correlations, without the need to invoke couplings
with an additional system (e.g. an environment). Indeed $\rho (t)$ cannot
operationally be distinguished from the density matrix $\rho ^{cc}(t)=\sum
p_{N}^{cl}\left\vert N\right\rangle \left\langle
N\right\vert \left\vert F(\bar{\epsilon}_{N})\right\rangle \left\langle F(%
\bar{\epsilon}_{N})\right\vert $ containing only classical correlations (and
for which $D(\rho ^{cc})=0$): the reduced density matrices obtained from $%
\rho $ and $\rho ^{cc}$ are identical, and as $\hslash \rightarrow 0$ the
coherences (in the \textquotedblleft pointer basis\textquotedblright
$\left\vert F(\bar{\epsilon}_{N})\right\rangle \left\vert N\right\rangle$) of
typical two-particle observables would lead to interference patterns with
vanishing (and therefore undetectable) wavelengths \cite{ballentine}.

To sum up, we have investigated entanglement evolution when the entanglement
is generated by a dynamical localized interaction in a quantum system having
a well defined classical counterpart. We have seen that entanglement
increases, irrespective of whether the underlying dynamics is regular or
chaotic, as typical actions grow relative to $\hbar $. The quantum
correlations are then given by the mutual information of the corresponding
classical system. The present results could contribute to a better
understanding of the role played by quantum information in the classical
limit.

\end{document}